\documentclass[
amssymb,aps,
prd,nofootinbib,floatfix,12pt,superscriptaddress,
]{revtex4}%
\usepackage[T1]{fontenc}
\usepackage[utf8]{inputenc}
\usepackage{textcomp}
\usepackage{amssymb}
\usepackage{amsmath}
\usepackage{epsfig}
\usepackage{graphicx}
\usepackage{esint}
\usepackage{comment}
\usepackage{xcolor}
\usepackage{color}
\PassOptionsToPackage{normalem}{ulem}
\usepackage{ulem}
\usepackage{slashed}
\usepackage{caption}
\usepackage{amsmath,amssymb,epsf}
\usepackage{graphicx,wrapfig,color}
\usepackage{subfig}
\usepackage{setspace}
\usepackage{makecell}
\usepackage{tabularx}
\usepackage{booktabs}
\usepackage[tablename=Table]{caption}
\usepackage{float}
\usepackage{placeins}
\usepackage[unicode=true,pdfusetitle,
 bookmarks=true,bookmarksnumbered=false,bookmarksopen=false,
 breaklinks=false,pdfborder={0 0 0},backref=false,colorlinks=true,citecolor=blue,urlcolor=violet 
]{hyperref}
\DeclareMathOperator{\sech}{sech}
\usepackage{xcolor}
\usepackage{footnote}

\newcommand{\orcid}[1]{\href{https://orcid.org/#1}{\includegraphics[width=10pt]{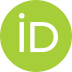}}}
\newcommand{\be}{\begin{equation}}
\newcommand{\ee}{\end{equation}}
\newcommand{\bea}{\begin{eqnarray}}
\newcommand{\eea}{\end{eqnarray}}

\newcommand{\ba}{\begin{array}}
\newcommand{\ea}{\end{array}}
\newcommand{\bd}{\begin{displaymath}}
\newcommand{\ed}{\end{displaymath}}

\def\gev{{\rm \,Ge\kern-0.125em V}}
\def\tev{{\rm \,Te\kern-0.125em V}}

\def\th13 {\theta_{13}}

\DeclareMathOperator{\csch}{csch}
\begin{document}
\title{A Case Study of Small Field Inflationary Dynamics in the Einstein-Gauss Bonnet Framework in the Light of $GW170817$}
\author{Mayukh R. Gangopadhyay\orcid{0000-0002-1466-8525}}
\email{mayukh_ccsp@sgtuniversity.org, mayukhraj@gmail.com}
\affiliation{Centre for Cosmology and Science Popularization, SGT University, Gurugram, Haryana-122505, India}

\author{Hussain Ahmed Khan\orcid{0000-0003-1166-5190}}
\email{hussaincyber@gmail.com}
\affiliation{Centre for Cosmology and Science Popularization, SGT University, Gurugram, Haryana-122505, India}

\author{Yogesh\orcid{0000-0002-7638-3082}}
\email{yogesh.ccsp@sgtuniversity.org, yogeshjjmi@gmail.com}
\affiliation{Centre for Cosmology and Science Popularization, SGT University, Gurugram, Haryana-122505, India}

\begin{abstract}
We study two of the most theoretically promising models of inflation, namely Natural inflation and Mutated Hilltop inflation, in the Einstein-Gauss Bonnet(EGB) gravity framework. In this work, we try to explore these models keeping observations from $GW170817$ on the speed of gravitational wave to be equal to the speed of light. This has direct implications on the non-minimal coupling to the Gauss-Bonnet invariant in the action. Thus, the effective potential gets new features. We have not only analysed the inflationary dynamics, but also the reheating dynamics and finally the corresponding energy spectrum of the gravitational wave.

\end{abstract}
\maketitle

\newpage

\section{Introduction}

In his seminal paper, Alan Guth\cite{guth} proposed the inflationary paradigm to overcome the several shortcomings that the standard hot big bang model faced\cite{Liddle,Linde:1981mu,Linde:1983gd,Linde:1982uu} (reviews can be found in \cite{Baumann:2009ds,cmbinflate,Linde:1984ir,mukha81,Linde:2007fr}). Since then it has become one of the primary fields of interest at the interface of cosmology and particle physics (readers are advised to go through \cite{Linde:2005ht,Lyth:1998xn} for important early works). Ever since the advancement in the observational capabilities in cosmology, it has become a part of precision physics. We can now use the observational data that can precisely guide us towards ever more accurate models of inflation, as it is  possible to restrict theoretical predictions using this data. Using CMB measurements from WMAP \cite{WMAP9} and \textit{Planck'18} \cite{PlanckXX}, most of the conventional inflationary models have been ruled out, at least in the cold inflationary scenario. The final data of \textit{Planck'18} \cite{Planck2018} also provides us with a rather fundamental insight, according to which, sub-Planckian small field models of inflation are favoured, which in a sense, is satisfying from the effective field theory perspective. 

The large field inflationary models have almost been ruled out by the Planck observations, and it is rather intriguing to note that small field inflationary models are favored, where the field value remains under the Planck mass. Indecently, this is also much easier to endorse from the effective field theory model building. 
In this work, we take two of the most elegant small-field models of inflation; Natural inflation\cite{freese} and Mutated Hilltop inflation\cite{Pal:2009sd}. It will be interesting as well as important to see if Natural inflation can survive in the EGB GW170817 compatible framework because in the standard scenario, it does not survive due to the severe observational constraints. Mutated Hilltop inflation is a well-motivated model as well, and it comes from the supergravity-inspired models, first proposed in \cite{Pal:2009sd}.

Natural inflation model was suggested \cite{freese} (also \cite{Adams:1992bn,Kim:2004rp,delaFuente:2014aca,Jensen:1986nf,Freese:2014nla}). This model employed an axion as the inflaton, which is the Goldstone  \cite{croon} of a spontaneously broken Peccei–Quinn symmetry. With a breaking scale of $10 M _{pl}$ or above, the model remains $2-\sigma$ consistent from the observational point of view. However, this has severe drawbacks since the dynamics of the effective field theory may be jeopardized by the effects of quantum gravity, which is expected to play a key role in the super-Planckian domain. In general, quantum gravity does not preserve global symmetry, hence having a super-Planckian breaking scale in the case of vanilla natural inflation, puts the whole thing theoretically, in unavoidable trouble.

When it comes to the Mutated Hilltop model, its beauty lies in the fact that it has the infinite number of terms in higher power, instead of the two-term approximation that is incorporated in many of the models of hilltop inflation \cite{Boubekeur:2005zm,Tzirakis:2007bf,Pal:2010eb,Stewart:1994pt}. This model, in the standard inflationary scenario, does survive the observational litmus test, but due to its origin in supergravity, it is rather interesting to check if it will survive in the framework we are considering. 

It has been observed that under modified theories of gravity, the limits on inflationary models may be made less rigorous \cite{nima,lindehyb,Koshelev:2020xby,nflation1,nflation2,Gerbino:2016sgw,nflation3,axionmono,pngb,Kallosh:2013pby,predictions,Barvinsky:1994hx,Cervantes-Cota1995,BezrukovShaposhnikov,EOPV2014}. In the sub-Planckian regime, we will investigate the validity of Natural inflation and Mutated Hilltop inflation, in the domain Einstein-Gauss-Bonnet gravity. The Gauss-Bonnet term is added to the Einstein-Hilbert action here, which has no effect on the equations of motion, as it is a total derivative. However, when coupled with a scalar field function $\phi$ as $\xi(\phi)$, it becomes dynamically important. There are several points of interest for working in the EGB framework. One is that in the EGB gravity, there are no extra degrees of freedom that we need to introduce, all the terms in the action, are either the functions of the scalar field or the curvature. Also, the Gauss-Bonnet term represents an intrinsic quantum correction to the Einstein-Hilbert action in the domain of string theory.

Taking the example of natural inflation, one of the theoretically well-motivated potentials, that we have studied here in EGB, is almost ousted by the latest Planck observations in the standard scenario. To resurrect this model, there have been several attempts in modified theories of gravity, such as braneworld scenario \cite{Neupane:2014vwa,Gangopadhyay:2016qqa,Calcagni:2013lya}. However the intrinsic nature of the braneworld theory pushes the value of the tensor to scalar ratio even to a higher value with respect to the standard scenario. This is in direct conflict with the stringent demand of $r \sim 0.059$ at the $1-\sigma$ level by the observations of Planck$'18$. The form of natural inflation can be motivated from different theoretical scenarios such as axiverse\cite{Cicoli:2012sz}, Chern-Simons background\cite{Bagherian:2022mau,Bhattacharjee:2014toa}, Goldstone inflation\cite{croon,Bhattacharya:2018xlw} etc. In this paper, we wanted to stick to the form of the potential originally proposed in \cite{freese}.

It has been shown that in the EGB framework, \cite{Soda2008,Guo:2009uk,Guo,Jiang:2013gza,Koh,vandeBruck:2015gjd,Pozdeeva:2020apf,Pozdeeva:2016cja,JoseMathew,vandeBruck:2016xvt,Koh:2016abf,Nozari:2017rta,Armaleo:2017lgr,Chakraborty:2018scm,Yi:2018gse,Odintsov:2018,Fomin:2019yls,Kleidis:2019ywv,Rashidi:2020wwg,Odintsov:2020sqy,Pozdeeva:2020shl,Kawai:2021bye,Kawai:2017kqt,Khan:2022odn,Kawai:2021edk,Nojiri:2006je,Kanti:2015pda,Kanti:1998jd,Vernov000,Pozdeeva:2021iwc,Oikonomou:2021kql,Ai:2020peo,Creminelli:2017sry,Ezquiaga:2017ekz,Hikmawan:2015rze}, the spectral index remains unchanged, and the tensor-to-scalar ratio becomes small, which is consistent with the observations. The most widely investigated models with GB coupling involve the function $\xi (\phi)$ being inversely proportional to
the scalar field potential~\cite{Guo:2009uk,Guo,Jiang:2013gza,Chakraborty:2018scm, Koh:2016abf,Yi:2018gse,Odintsov:2018,Kleidis:2019ywv,Rashidi:2020wwg,Pozdeeva:2020shl}. In view of the observations from GW170817, there is constraint on the speed of the gravitational wave propagation. The speed of propagation has to be equal to the speed of light, which, in natural units is equal to unity ($c_{T}=1$). This new constraint gives us a differential equation relating the scalar potential considered and the type of EGB coupling $\xi (\phi)$ we can use, for the particular scalar potential that we are considering. It leads us to the fact that we cannot pick any random coupling $\xi (\phi)$.

We will also analyze the type of gravitational energy spectrum that is produced by the two considered inflationary models, and the fundamental role that the value of the tensor spectral index $n_{T}$ plays in the GW spectrum enhancement, or lack thereof. 

The remainder of the paper is structured as follows. In section \ref{sec2} we will go over some of the fundamental but critical features in the EGB framework. In section In section \ref{effectivepotential} we go over the effective potential approach. In \ref{sec_GW} we draw the basics of the primordial GW calculations.  In section \ref{small_field} we discuss the two models which we have considered for the inflationary analysis. Dynamics of reheating and the primordial gravitational wave energy spectrum can be found in the section \ref{reheatingegb} and \ref{GWANA} respectively. All the results of our analysis of inflation, reheating and GW energy spectrum can be found in section \ref{RES}. Finally, in \ref{conslusion} we draw the conclusion of our analyses.

\subsection{Review of Einstein-Gauss Bonnet Gravity}
\label{sec2}

In the following analyses we will be considering a modified gravity model with the invariant GB term,~\cite{Pozdeeva:2020apf,Guo:2009uk,Guo}:

\begin{equation}
\label{action1}
S=\int d^4x\sqrt{-g}\left[UR-\frac{g^{\mu\nu}}{2}\partial_\mu\phi\partial_\nu\phi-V(\phi)-\frac{\xi(\phi)}{2}\mathcal{G}\right],
\end{equation}
here, $V(\phi)$, and $\xi(\phi)$ are differentiable functions, $U$ is a positive constant, and $\mathcal{G}$ is the 4-dimensional Gauss-Bonnet term given as, $$\mathcal{G}=R_{\mu\nu\rho\sigma}R^{\mu\nu\rho\sigma}-4R_{\mu\nu}R^{\mu\nu}+R^2$$

Varying the action (\ref{action1}) with respect to the scalar field $\phi$, gives us the following set of equations, in a spatially flat Friedmann Universe, ~\cite{vandeBruck:2015gjd,Pozdeeva:2019agu}: 

\begin{eqnarray}
               12UH^2 &=& \dot\phi^2+2V+24\dot{\xi}H^3,  \label{Equ0} \\
                4U\dot{H} &=& {}-\dot\phi^2+4\ddot{\xi}H^2+4\dot{\xi}H\left(2\dot{H}-H^2\right), \label{EquH}\\
                \ddot{\phi}&=&{}-3H\dot{\phi}-V'-12\xi'H^2\left(\dot{H}+H^2\right), \label{Equphi}
\end{eqnarray}
here, primes and dots represent the derivatives taken with respect to the  the scalar field  $\phi$ and cosmic time $t$ respectively, and $H=\dot{a}/a$ is the Hubble parameter, where $a$ is the scale factor.

Using the slow-roll parameters as given in Refs.~\cite{Guo,vandeBruck:2015gjd}:
\begin{eqnarray}
  \epsilon_1 &=&{}-\frac{\dot{H}}{H^2}={}-\frac{d\ln(H)}{dN},\qquad \epsilon_{i+1}= \frac{d\ln|\epsilon_i|}{dN},\quad i\geqslant 1, \\
  \delta_1&=& \frac{2}{U}H\dot{\xi}=\frac{2}{U}H^2\xi'\frac{d{\phi}}{dN},\qquad \delta_{i+1}=\frac{d\ln|\delta_i|}{dN},\quad i\geqslant 1,
\end{eqnarray}
here we are considering ${d}/{dt}=H\, {d}/{dN}$. 

The slow-roll approximation then requires, $$ |\epsilon_i| \ll 1 \; \text{and} \;  |\delta_i| \ll 1  
$$

We simplify Eqs.~(\ref{Equ0})--(\ref{Equphi}), using slow-roll conditions $\epsilon_1 \ll 1, \; \epsilon_2 \ll 1 \; \delta_1 \ll 1 \; \text{and} \; \delta_2 \ll 1. $

Hence, we get,
\begin{equation}
\label{delta2equ}
\delta_2=\frac{\dot\delta_1}{H\delta_1}=\frac{2\ddot\xi}{U\delta_1}-\epsilon_1,
\end{equation}
and we have $|\ddot\xi|\ll |H\dot\xi|$ from $|\delta_2|\ll 1$ and $|\epsilon_1|\ll 1$.

Using the conditions $|\delta_1|\ll 1$ and $|\delta_2|\ll 1$, in Eqs.~(\ref{Equ0}) and (\ref{EquH}), we get:
\begin{eqnarray}
               &&12UH^2 \simeq \dot\phi^2+2V,  \label{Equ0slr1} \\
                &&4U\dot{H} \simeq{}-\dot\phi^2-4\dot{\xi}H^3={}-\dot{\phi}\left(\dot{\phi}+4\xi'H^3\right). \label{EquHslr1}
\end{eqnarray}
Using,
\begin{equation*}
\epsilon_1={}-\frac{\dot H}{H^2}\simeq \frac{\dot\phi^2}{3(\dot\phi^2+2V)}+\frac{1}{2}\delta_1\ll 1,
\end{equation*}
we obtain $\dot\phi^2 \ll 2V$, and Eq.~(\ref{Equ0slr1}) takes the following form
\begin{equation}
\label{Equ0slr2}
6UH^2 \simeq V.
\end{equation}
Differentiating the above equation with respect to time and making use of Eq.~(\ref{EquHslr1}), we obtain
\begin{equation}
\label{Equphislr1}
    \dot{\phi}\simeq{}-\frac{V'}{3H}-4\xi'H^3.
\end{equation}
Substituting (\ref{Equphislr1}) into Eq.~(\ref{Equphi}), we get $|\ddot{\phi}|\simeq|12\xi'H^2\dot{H}|\ll|12\xi'H^4| $.

Therefore, the slow-roll conditions give the following results:
 \begin{equation*}
 \dot\phi^2\ll V, \quad |\ddot{\phi}|\ll |12\xi'H^4|,\quad 2|\dot{\xi}|H\ll U,\quad |\ddot{\xi}|\ll|\dot{\xi}|H\,,
 \end{equation*}
 and the leading order equations in the slow-roll approximation will be:
 \begin{eqnarray}
               H^2&\simeq&\frac{V}{6U}\,, \label{Equ0lo}\\
               \dot{H}&\simeq&{}-\frac{\dot\phi^2}{4U}-\frac{\dot{\xi}H^3}{U}\,, \label{EquHlo}\\
                \dot{\phi}&\simeq&{}-\frac{V'+12\xi'H^4}{3H}. \label{Equphilo}
\end{eqnarray}

\subsection{The effective potential}
\label{effectivepotential}

The stability of de Sitter solutions for the model (\ref{action1}) is analyzed by making use of the effective potential approach, introduced in Ref.~\cite{Pozdeeva:2019agu}:
\begin{equation}
\label{Veff}
V_{eff}(\phi)={}-\frac{U^2}{V(\phi)}+\frac{1}{3}\xi(\phi).
\end{equation}

We employ the effective potential approach in order to study the stability of de Sitter solutions in a model with a non-minimally coupled scalar field. The zeros of the first derivative of the effective potential then give us the de Sitter points, and the sign of the  second derivative informs us of the stability of the de Sitter solution. The effective potential approach is especially ingenious to compare the de Sitter solutions without the GB term (for further details, the reader is suggested to go through \cite{Pozdeeva:2019agu}).

Inflationary scenarios will be unstable for the case, $U= \frac{1}{2}$, as  $V(\phi)$ becomes undefined. ~\cite{Hikmawan:2015rze} (see also~\cite{Chakraborty:2018scm}).

In our work, the inflationary scenarios that we will be considering will be with positive potentials only: $V(\phi)> 0$ during inflation.
The existence and stability of de Sitter solutions can be characterized completely by the effective potential. 
But we keep in mind that, as the potential $V(\phi)$ enters in the equations, to get inflationary parameters, the effective potential is not enough to characterize completely the quasi-de Sitter inflationary stage. 
However we keep the effective potential in the corresponding formulae, as it proves to be useful in the analysis. 

Using Eqs.~(\ref{EquHlo}) and (\ref{Equphilo}), we get that the functions $H(N)$ and $\phi(N)$ satisfy the following leading order equations:
\begin{eqnarray}
               \frac{d{H}}{dN}&\simeq&{}-\frac{H}{U}V'V_{eff}'\,, \label{EquHloN}\\
               \frac{d{\phi}}{dN}&\simeq&{}-2\frac{V}{U}V_{eff}'. \label{EquphiloN}
\end{eqnarray}

The slow-roll parameters, in terms of the effective potential, become:

\begin{equation}
    \epsilon_1={}-\frac{1}{2}\frac{d\ln(V)}{dN}=\frac{V'}{U}V_{eff}'\,,
    \label{EGBsr1}
    \end{equation}
\begin{equation}
    \epsilon_2={}-\frac{2V}{U}V_{eff}'\left[\frac{V''}{V'}+\frac{V_{eff}''}{V_{eff}'}\right]
    ={}-\frac{2V}{U}V_{eff}'\left[\ln(V'V_{eff}')\right]'\,,
    \label{EGBsr2}
\end{equation}
\begin{equation}
  \delta_1= {}-\frac{2V^2}{3U^3}\xi'V_{eff}'\,,
  \label{EGBsr3}
  \end{equation}
\begin{equation}
\begin{split}
\delta_2=& {}-\frac{2V}{U}V_{eff}'\left[2\frac{V'}{V}+\frac{V_{eff}''}{V_{eff}'}+\frac{\xi''}{\xi'}\right]\\
=&{}-\frac{2V}{U}V_{eff}'\left[\ln(V^2\xi'V_{eff}')\right]'.
\end{split}
\label{slrVeffd}
\end{equation}

Therefore, $|\epsilon_1|\ll 1$ and $|\delta_1|\ll 1$ given $V_{eff}'$ is small enough. This will enable us to construct inflationary scenarios in models with the GB term, using the effective potential. 

For the tensor-to-scalar ratio ($r$), the scalar spectrum index ($n_{s}$), and the tensor spectrum index ($n_{T}$), we use the known formulae~\cite{Guo,Koh:2016abf}, and get,

\begin{equation}
\label{rVeff}
 r=8|2\epsilon_1-\delta_1|
\end{equation}
\begin{equation}
\label{nsVeff}
\begin{split}
   n_s=&1-2\epsilon_1-\frac{2\epsilon_1\epsilon_2-\delta_1\delta_2}{2\epsilon_1-\delta_1}
\end{split}
\end{equation}
\begin{equation}
\label{ntVeff}
    n_{T} = -\frac{2\epsilon_{1}}{1-\epsilon_{1}}
\end{equation}

A standard way to reconstruct inflationary models~\cite{Mukhanov:2013tua,Koh:2016abf,Pozdeeva:2020shl} assumes the explicit form of the inflationary parameter $n_s$ and $r$ as functions of $N$.  

The amplitude $A_{s}$, to a leading order of approximation, is given by the expression~\cite{vandeBruck:2015gjd},

\begin{equation}
\label{As}
A_s\approx\frac{H^2}{\pi^2 U r}\approx\frac{V}{6\pi^2 U^2r}.
\end{equation}

The e-folding number $N$, in the slow-roll approximation, can be represented as a function of $\phi$ as:

\begin{equation}
\label{N1}
N(\phi)=\int\limits^{\phi}_{\phi_{end}}\frac{U}{2VV'_{eff}}d\phi.
\end{equation}

To obtain viable inflationary scenarios, the calculations of the inflationary parameters is done between the $50 \leq N \leq 70$ e-foldings.

\subsection{GW170817 Compatible EGB Models}
\label{sec_GW}
Now in order make the Einstein-Gauss-Bonnet models compatible with the observations from GW170817, we need to put constraints on the speed of the tensor perturbations\cite{Oikonomou:2022xoq,LIGOScientific:2017vwq,Oikonomou:2022ksx,Nojiri:2022ski} (readers are suggested to go trough \cite{reviews3,congola000, Odintsov:2020xji,Odintsov:2020mkz,Oikonomou:2020sij,Giare:2020vss,Zhang:2021ygh}) The speed of tensor perturbations has to be equal to the speed of light\cite{LIGOScientific:2017vwq}.

The expression for the speed of tensor perturbations reads as,
\begin{equation}
\label{CT1}
    c_{T}^{2} = 1-\frac{Q_{f}}{2Q_{t}}
\end{equation}
Here the functions $Q_{f}$, $Q_{t}$ and $F$ are defined as,
\begin{equation}
    Q_{f}=8(\ddot{\xi}-h\dot{\xi}),\;\;\; Q_{t}=F+\frac{Qb}{2},\;\;\; F=\frac{1}{\kappa^{2}}\;\; \text{and} \;\; Q_{b}=8\dot{\xi}H
\end{equation}

To make $c_{T}^{2} = 1$, the function $Q_{f}$ must vanish. 
\begin{equation}
    Q_{f} = 8(\ddot{\xi}-H\dot{\xi})=0
\end{equation}
\begin{equation}
    \implies \ddot{\xi} = H\dot{\xi}
\end{equation}

It is seen that the imposition of the condition on the speed of tensor perturbations gives us a differential equation that constrains the coupling function $\xi (\phi)$. 

In terms of the scalar field, this equation reads as,
\begin{equation}
\label{constrEQ}
    \xi^{\prime\prime} \dot{\phi^2} + \xi^{\prime} \ddot{\phi} = H \xi^{\prime} \dot{\phi}
\end{equation}

If we assume $\xi^{\prime} \ddot{\phi} \ll \xi^{\prime\prime} \dot{\phi^{2}}$, then (\ref{constrEQ}) gives,
\begin{equation}
\label{phicondn}
    \dot{\phi} \simeq \frac{H \xi^{\prime}}{\xi^{\prime\prime}}
\end{equation}

Combining the equation above with (\ref{Equphi}), we get,
\begin{equation}
\label{XIHV}
    \frac{\xi^{\prime}}{\xi^{\prime\prime}} \simeq - \frac{1}{3H^{2}} (V^{\prime} + 12 \xi^{\prime} H^{4})
\end{equation}

We will further impose the followings conditions,

\begin{equation}
    \kappa \frac{\xi^{\prime}}{\xi^{\prime\prime}} \ll 1 \;\; \text{and} \;\; 12\frac{\xi^{\prime 2} H^{4}}{\xi^{\prime\prime}} \ll V
\end{equation}

Now in view of the above conditions and relations, the differential equation (\ref{XIHV}) becomes,
\begin{equation}
\label{MAINDFE}
    \frac{V^{\prime}}{V^{2}} + \frac{4\kappa^{2}}{3} \xi^{\prime} \simeq 0
\end{equation}

This equation will prove to be quintessential in our analysis as we will use this differential equation to solve for the scalar coupling $\xi (\phi)$, with the particular inflationary potential as the input. 

 
\subsection{Small field models of inflation}
In this section we will explore two small field models of inflation namely Natural and Mutated Hilltop and their dynamics. 
\label{small_field}
\begin{center}
\bf{Natural Inflation} 
\end{center}

Natural inflation was first introduced in \cite{freese}. There is ample literature showing that in the cold inflationary scenario, with sub-Plackian values of the breaking scale $f$, it is not possible to resurrect Natural inflation with the current observational constrains. So, it will be interesting to examine the validity of Natural inflation within the context of EGB gravity with the constraint on the coupling scalar function $\xi(\phi)$ given by the $c_{T}=1$ condition. Natural inflation potential can be written as,
\begin{equation}
V = V_0 \bigg [1+ \cos \bigg(\frac{\phi}{f}   \bigg) \bigg].
\label{natural_potential}
\end{equation}

For the above given form of the potential, using Eqn. (\ref{MAINDFE}), we can work out the scalar coupling function, which turns out to be, $\xi(\phi) =  \frac{3 \sec^2 \left(\frac{\phi }{2 f}\right)}{8 \Lambda}+c_1  $, where $c_1$ is a constant of integration. However, this constant can be ignored altogether in our analyses for the fact that it cancels out during the calculations of the slow-roll parameters. For Natural Inflation, the slow-roll parameters can be expressed as following:

 
 \begin{equation}
     \epsilon_1 = \frac{(-1+4 U^2) \tan^2 \left(\frac{\phi}{2 f }  \right)}{4 f^2 U}, ~~~~ \epsilon_2 = \frac{(-1+4 U^2) \sec^2 \left(\frac{\phi}{2 f }  \right)}{2 f^2 U}
     \label{slow_natural_1}
 \end{equation}
 similarly, 
 \begin{equation}
   \delta_1 = \frac{(-1+4 U^2) \tan^2 \left(\frac{\phi}{2 f }  \right)}{8 f^2 U^3}, ~~~~\delta_2= \frac{-1+4 U^2}{ f^2 U \left(1+ \cos \left( \frac{\phi}{f}  \right) \right)}
   \label{slow_natural_2}
 \end{equation}
The set of slow roll parameters (\ref{slow_natural_1}) and (\ref{slow_natural_2}), allow us write the tensor-to-scalar ratio ($r$), scalar spectral index ($n_s$) and tensor spectral index ($n_T$) in the following form:

\begin{equation}
r = \Bigg |\frac{(1-4 U^2)^2 \tan^2 \left(\frac{\phi}{2 f }  \right)}{f^2 U^3} \Bigg|,~~~ n_s = \frac{-1+2~U(f^2+ 2U )+ (2-8~U^2) \sec^2 \left(\frac{\phi}{2 f}  \right)}{2 f^2~U} 
\end{equation}
\begin{equation}
n_T = 2 - \frac{8 f^2 U}{4 f^2 U+ \left(1-4 U^2 \right) \tan^2 \left(\frac{\phi}{2 f} \right)}
\end{equation}
From the Eq. (\ref{N1}), we can compute the number of e-folds during the inflationary epoch
\begin{equation}
N(\phi)=\int\limits^{\phi_i}_{\phi_{end}} \frac{2 f U \cot{\left( \frac{\phi}{2 f} \right)}}{1-4U^2}
\label{efold_natural}
\end{equation}
where, $\phi_i$ and $\phi_{end}$ are the field values at the time of horizon exit and end of inflation respectively. Field value at the end of inflation can be evaluated by imposing the condition $\epsilon_1=1$, substituting this in the Eq. (\ref{efold_natural}) we can write filed value at the time of horizon exit in terms of number of e-folds as $\phi_i = 2 f \sin^{-1} \left( e^{\frac{N-4NU^2}{4 f^2 U}} \right)$. This allows us to calculate the inflationary observables $r$ and $n_s$. The scale of inflation ($V_0$) can be fixed from the Eq. (\ref{As})

\begin{center}
\bf{Mutated Hilltop Inflation} 
\end{center}

This potential belongs to a particular class of models, called the Hilltop potentials \cite{Boubekeur:2005zm,Tzirakis:2007bf}. Inflation is expected to occur at the peak of the potential in such class of models. It was demonstrated in Refs.\cite{Boubekeur:2005zm,Tzirakis:2007bf}, that by adding contributions from higher-order operators, F or D term inflation may be converted into hilltop models. In this section, we analyze the Mutated hilltop inflation, which was initially described, and explored in references \cite{Pal:2009sd,Pal:2010eb,Stewart:1994pt}. The potential is purely phenomenological, the form of which is given as,
 
\begin{equation}
V = V_0 \bigg [1 -\sech \big( \alpha \phi \big) \bigg].
\label{mutated_potential}
\end{equation}

Eqn.(\ref{MAINDFE}) allows us to calculate the EGB coupling for the potential (\ref{mutated_potential}), which leads to the following $\xi(\phi) =\frac{3 \csch^2 \left(\frac{\alpha \phi }{2 }\right)}{8 \Lambda} +c_1$, where $c_1$ is again a constant of integration that can be ignored for the aforementioned reasons. For the potential in consideration, the slow-roll parameters take the following form:
\begin{align}
&\epsilon_1 = \frac{(-1 +4 U^2) \alpha^2 \coth^2{\left(\frac{\alpha \phi}{2}  \right)} \sech^2{\left (\alpha \phi \right)}}{4 U}\nonumber \\
&\delta_1 = \frac{2 \left(  -1+4 U^2\right) \alpha^2 \cosh^4{\left(\frac{\alpha \phi}{2}  \right)} \csch^2{\left( 2 \alpha \phi\right)}}{U^3}\nonumber \\
&\epsilon_2 =\delta_2= \frac{ \left(-1+4 U^2\right) \alpha^2 \left(-1+2 \cosh{ \left(\alpha \phi\right)}+ \cosh{\left(2 \alpha \phi\right)} \right) \csch^2{\left(\frac{\alpha \phi}{2}\right) \sech^2{ \left( \alpha \phi \right)}}}{4 U}
\label{slow_mutated_1}
\end{align}
Inflationary observables, $r$ and $n_s$, can be computed with the help of slow-roll expressions (\ref{slow_mutated_1}).The tensor-to-scalar ratio ($r$), scalar spectral index ($n_s$), and the tensor spectral index ($n_T$), are thus expressed as:

\begin{equation}
r= \Bigg| \frac{\left( 1-4 U^2 \right)^2 \alpha^2 \coth^2{\left( \frac{\alpha \phi}{2} \right)} \sech^2{\left(   \alpha \phi \right)}}{U^3}\Bigg|,
\label{r_mutated}
\end{equation}
\begin{equation}
n_s = 1 - \frac{\left(-1+4 U^2\right) \alpha^2 \left( 2 \csch^2{\left(\frac{\alpha \phi}{2}\right)+ \left(-2+ \sech{ \alpha \phi}\right)}~\sech{ \alpha \phi}\right)}{2 U}
\label{ns_mutated}
\end{equation}
\begin{equation}
n_T = 2 + \frac{8 U}{-4U+ \left(-1+4U^2\right) \alpha^2  \coth^2{\left(\frac{\alpha \phi}{2} \right)} \sech^2{\alpha \phi}} 
\label{nt_mutated}
\end{equation}

For the number of e-folds, we again make use of Eqn.(\ref{N1}), and arrive at the following expression:
 \begin{equation}
N(\phi)=\int\limits^{\phi_i}_{\phi_{end}} -\frac{2 U \cosh{\left(\alpha \phi  \right)} \tanh{\left( \frac{\alpha \phi}{2} \right)} }{\alpha- 4 U^2 \alpha}
\label{efold_mutated}
\end{equation}
where, $\phi_i$ and $\phi_{end}$ are the inflaton field values, at the time of horizon exit, and at the end of inflation respectively. Field value at the end of inflation can be evaluated by imposing the condition $\epsilon_1=1$. However, an analytical approach towards solving Eqn. (\ref{efold_mutated}) is difficult, so we use numerical techniques to solve for the field value at the time of horizon exit in terms of numbers of e-folds. The scale of inflation ($V_0$) can be fixed from the Eq. (\ref{As})

\subsection{Dynamics of Reheating }
\label{reheatingegb}
As the universe grows exponentially, it ends up in a super-cooled state after inflation. So, for the universe to reheat itself, enter the radiation-dominated epoch, and begin the Big Bang Nucleosynthesis (BBN), a mechanism for the universe to exit this super-cooled state is required. \cite{33, 34, 35, 36, 37, 38, 39} (for reheating in EGB readers are advised to go through\cite{Kohreheating,vandeBruck:2016xvt,Nozari:2017rta,Odintsov:2022sdk}). Other realisations of inflationary dynamics in nonstandard settings, such as warm inflation, are proposed to readers in ref. \cite{40, 41, 42, 43,44,44a}, where the reheating phase is not necessary, we proceed immediately into the radiation-dominated phase following the end of inflation.
 This evolution of the universe from a super-cooled state to a hot, thermal, and radiation-dominated state can occur via a perturbative process known as perturbative reheating or a parametric resonance process (p)reheating. (The readers are advised to go to \cite{45} for a more extensive discussion.) The reheating period may be parameterized by $N_{re}$ (the number of e-folds during the reheating phase), $T_{re}$ (the thermalization temperature), and the equation of states during reheating ($\omega_{re}$) \cite{46, 47}. The approach we adopt does not require the precise dynamical reheating process, yet one may still explore the parameter space for   reheating. \cite{48,Adhikari:2019uaw}. 

\begin{equation}
N_{re}= \frac{4}{ (1-3w_{re} )}   \left[61.488  - \ln \left(\frac{ V_{end}^{\frac{1}{4}}}{ H_{k} } \right)  - N_{k}   \right]
\label{re7}
\end{equation}

\begin{equation}
T_{re}= \left[ \left(\frac{43}{11 g_{re}} \right)^{\frac{1}{3}}    \frac{a_0 T_0}{k_{}} H_{k} e^{- N_{k}} \left[\frac{3^2 \cdot 5 V_{end}}{\pi^2 g_{re}} \right]^{- \frac{1}{3(1 + w_{re})}}  \right]^{\frac{3(1+ w_{re})}{3 w_{re} -1}}.
\label{re8}
\end{equation}

In this case, we employed Planck's pivot ($k$) of order $0.05 \; \mbox{Mpc}^{-1}$ and $g_{re} \approx 106 $ (for $\mbox{SUSY},~g_{re}\approx 226 $ ). To assess $N_{re}$ and $T_{re}$, one must first compute $H_k$, $N_k$, and $V_{end}$, for the given potential. We can establish the relationship between $H_k$ and $n_s$ using the Eq. (\ref{As}), and similarly, $N_k$ may be expressed in terms of the spectral index ($n_s$). It is trivial to determine $V_{end}$ from the $\epsilon=1$ condition at the end of inflation. We compute the reheating temperature and number of e-folds during reheating for the different equations of state ($\omega_{re}$) once we have completed all of the preliminary calculations. We can estimate the reheating temperature and corresponding numbers of e-folding for the given potential.  
 
\newpage

\subsection{Primordial Gravitational Waves in Einstein Gravity}
\label{GWANA}

In this section we motivate the expression that we will be using to calculate the GW energy spectrum for the models that we have considered above, and how the spectrum is affected by the corrections in the EGB models in the view of GW170817. 

The energy density of gravitational waves is given by,
\begin{equation}
\rho_{GW} = \frac{1}{64 \pi G a^{2}} \langle (\partial h_{ij})^{2} + (\Delta h_{ij})^2 \rangle
\end{equation}
As each mode is quantified by the wavenumber $k$, we need to Fourier transform the tensor perturbations as, 
\begin{equation}
h_{ij} (\tau,\textbf{x}) = \sum_{\lambda = + \times} \int \frac{dk^{3}}{(2 \pi)^{\frac{3}{2}}} \epsilon^{\lambda}_{ij} h_{\textbf{k}}(\tau) e^{i \textbf{k.x}}
\end{equation}

The energy density of gravitational waves, $\rho_{GW}$ can be rewritten as,
\begin{equation}
\label{rhoGW}
    \rho_{GW} = \frac{1}{32 \pi G} \int d\;\text{ln}\;k \left(\frac{k}{a}\right)^{2} \frac{k^{3}}{\pi^{2}} \sum_{\lambda} |h_{\textbf{k}}^{\lambda}|^{2}.
\end{equation}

For a gravitational wave with wavenumber $k$, its amplitude remains constant for as long as the mode is lying outside the horizon. As soon as it renters the horizon, damping of the amplitude starts taking place.
For a mode entering during the matter dominated era we have \cite{Boyle:2005se,Odintsov:2021kup,Liu:2015psa,Zhao:2013bba,Nishizawa:2017nef,Arai:2017hxj,Nunes:2018zot}(see also \cite{Kawai:1998ab,Kawai:1999pw,Soda:1998tr} ),

\begin{equation}
    h_{k} (\tau) = h_{k}^{(p)} \left (\frac{3 j_{1}(k\tau)}{k\tau}\right)
\end{equation}
where $j_{i} (k\tau)$ is the Spherical Bessel Function of the first kind given by,

\begin{equation}
    j_{1} (k\tau) = \frac{\text{sin}(k\tau)}{(k\tau)^{2}}-\frac{\text{cos}(k\tau)}{k\tau}
\end{equation}.
One of the damping factors is due to the acceleration of the universe \cite{Boyle:2005se}
\begin{equation*}
    \left (\frac{\Omega_{m}}{\Omega_{\Lambda}}\right)^{2}
\end{equation*}
And the last damping factor that we need to address is due to the fact that the relativistic degrees of freedom, in the early universe, do not remain constant. This gives us a damping factor as, \cite{Odintsov:2021kup}
\begin{equation}
    \left( \frac{g_{*}(T_{in})}{g_{*0}} \right) \left (\frac{g_{*s0}}{g_{*s}(T_{in})} \right)^{4/3}.
\end{equation}

For this damping factor to be incorporated in the gravitational wave spectrum, we use the following fitting functions \cite{Kuroyanagi:2014nba}

\begin{equation*}
    g_{*}(T_{in} (k))=g_{*0} \left[ \frac{A+\text{tanh}\left[-2.5\text{log}_{10} \left(\frac{k/2 \pi}{2.5 \times 10^{-12} \text{Hz}}\right)\right]}{A+1}\right] \left[ \frac{B+\text{tanh}\left[-2\text{log}_{10} \left(\frac{k/2 \pi}{6.0 \times 10^{-12} \text{Hz}}\right)\right]}{B+1}\right]
\end{equation*}
where $A = \frac{-1-10.75/g_{*0}}{-1+10.75/g_{*0}}$ and $B = \frac{-1-g_{max}/10.75}{-1+g_{max}/10.75} $. Here $g_{*0}=3.36$, $g_{max}=106.75$, and $g_{*s}=3.91$.   

The present day gravitational energy wave spectrum is characterized in the form of energy density parameter per logarithmic interval of the wave number $k$,

\begin{equation}
    \Omega_{GW} (k) = \frac{1}{\rho_{crit}} \frac{d\rho_{GW}}{d\;\text{ln}\;k}
\end{equation}
where $\rho_{crit}=3H^{2}/(8\pi G)$.

We can use (\ref{rhoGW}), and get the final expression as,
\begin{equation}
\label{omegaGW}
    \Omega_{GW} = \frac{k^{2}}{12H_{0}^{2}} \Delta_{h}^{2}(k)
\end{equation}

where we have defined the power spectrum $\Delta_{h}^{2}(k)$ as,
\begin{equation}
\label{ps}
    \Delta_{h}^{2}(k) = T_{T}^{2}(k) \Delta_{h}^{(p)}(k)^{2}
\end{equation}

Here, $T_{T}(k)$ is the transfer function, which describes the evolution of the gravitational waves after horizon crossing. It has the following form \cite{Boyle:2005se,Liu:2015psa,Zhao:2013bba,Nishizawa:2017nef,Arai:2017hxj,Nunes:2018zot},
\begin{equation}
\label{TRF}
    T_{T}^{2} (k) =  \left (\frac{\Omega_{m}}{\Omega_{\Lambda}}\right)^{2}  \left( \frac{g_{*}(T_{in})}{g_{*0}} \right) \left (\frac{g_{*s0}}{g_{*s}(T_{in})} \right)^{4/3} \left(\frac{3 j_{1}(k\tau_{0})}{k\tau_{0}}\right)^{2} T_{1}^{2}(x_{eq}) T^{2}_{2} (x_{R})
\end{equation}

$ \Delta_{h}^{(p)}(k)^{2}$ is defined as the primordial power spectrum and it corresponds to the inflationary era, given as,
\begin{equation}
\label{PPS0}
    \Delta_{h}^{(p)}(k)^{2} = \mathcal{A}_{T}(k_{ref}) \left( \frac{k}{k_{ref}}\right)^{n_{T}}.
\end{equation}

Here $\mathcal{A}_{T}(k_{ref})$ is the amplitude at the reference scale $k_{ref}$ and $n_{T}$ is the spectral index. 

The amplitude $\mathcal{A}_{T}(k_{ref})$ is related to the scalar perturbation amplitude through the tensor-to-scalar ratio as,

\begin{equation}
    \mathcal{A}_{T}(k_{ref}) = r(k_{ref}) \mathcal{P}_{\zeta} (k_{ref})
\end{equation}

This relation can be used in (\ref{PPS0}) to obtain the relation for primordial tensor spectrum as, 

\begin{equation}
    \Delta_{h}^{(p)}(k)^{2} = r(k_{ref}) \mathcal{P}_{\zeta} (k_{ref}) \left( \frac{k}{k_{ref}}\right)^{n_{T}}.
\end{equation}

As for the two more transfer functions defined in (\ref{TRF}), we can write them as,\cite{Kuroyanagi:2014nba}
\begin{equation}
T_{1}^{2} (x_{eq}) = [1+1.57x_{eq}+3.42x_{eq}^{2}]\\
\end{equation}
\begin{equation}
T_{2}^{2} (x_{R}) = (1-0.22x_{R}^{1.5} +0.65x_{R}^{2})^{-1}
\end{equation}

with $x_{eq}=k/k_{eq}$ and $x_{R} = k/k_{R}$. The wavenumbers $k_{eq}$ and $k_{R}$ correspond to modes re-entering the horizon arond matter-radiation equality and modes re-entering after the phase of reheating has started, respectively. They are given as,
\begin{equation}
    k_{eq} = 7.1 \times 10^{-2} \Omega_{m} h^{2} \text{Mpc}^{-1}
\end{equation}
\begin{equation}
    k_{R} = 1.7 \times 10^{14} \left( \frac{g_{*s}(T_{R})}{106.75} \right)^{1/6}\left( \frac{T_{R}}{10^{7}\text{GeV} } \right) \text{Mpc}^{-1}
\end{equation}

Therefore we can write the final expression for the power spectrum as, 

\begin{eqnarray}
\label{OMEGAGW12}
       \Omega_{GW}(k) = \frac{k^{2}}{12H_{0}^{2}}r(k_{ref}) \mathcal{P}_{\zeta} (k_{ref}) \left( \frac{k}{k_{ref}}\right)^{n_{T}}  \left (\frac{\Omega_{m}}{\Omega_{\Lambda}}\right)^{2}  \left( \frac{g_{*}(T_{in})}{g_{*0}} \right)\nonumber \\ \left (\frac{g_{*s0}}{g_{*s}(T_{in})} \right)^{4/3} \left(\frac{3 j_{1}(k\tau_{0})}{k\tau_{0}}\right)^{2} T_{1}^{2}(x_{eq}) T^{2}_{2} (x_{R})          
\end{eqnarray}

\subsection*{Primordial Gravitational Waves in Einstein-Gauss-Bonnet Gravity}
In Einstein-Gauss-Bonnet gravity, the Fourier transform of the tensor perturbation of a flat FRW metric is governed by,\cite{Hwang:2005hb,Odintsov:2021kup}
\begin{equation}
\label{GWF}
\frac{1}{a^{3} Q_{t}} \frac{d}{dt} (a^{3} Q_{t} \dot{h}(k)) + \frac{k^{2}}{a^{2}} h(k) = 0.
\end{equation}
For Einstein-Gauss-Bonnet gravity, $Q_{t}$ is defined as 
\begin{equation}
Q_{t} = F + \frac{1}{\kappa^{2}}\;\;\;\;\;\;\;\; \text{where,} \;\; F=\frac{1}{\kappa^{2}}\;\;\; \text{and}\;\;\; Q_{b} = -8 \dot{\xi} H
\end{equation}
Now the evolution equation (\ref{GWF}) can be written as \cite{Oikonomou:2022xoq} ,
\begin{equation}
\label{hk}
\ddot{h} (k) + (3 + a_{M})H \dot{h}(k) +\frac{k^{2}}{a^{2}} h(k) = 0 
\end{equation}
where $a_{M}$ is,
\begin{equation}
    a_{M} = \frac{-4 \ddot{\xi}H - 4\dot{\xi}H}{H(\frac{1}{\kappa^{2}}-4 \dot{\xi}H)}.
\end{equation}

The gravitational wave waveform should change when we are working the framework of Einstein-Gauss-Bonnet gravity, instead of the standard theory of General Relativity. This is where the role of $a_{m}$ comes in, it represents the deviation of the waveform in EGB gravity.\newline

Now to solve the differential equation (\ref{hk}), a WKB method, developed in \cite{Nishizawa:2017nef,Arai:2017hxj} is used, according to which the solution of the equation is given as, 

\begin{equation}
    h_{EGB}=\text{e}^{\mathcal{-D}}h_{GR}
\end{equation}
 
The parameter $\mathcal{D}$ is written as an integral of the parameter $a_{M}$ and $\mathcal{H}$ over conformal time,

\begin{equation}
    \mathcal{D} = \frac{1}{2} \int^{\tau} a_{M}\mathcal{H}d\tau_{*}
\end{equation}

Using the relation $\frac{dz}{d\tau}=-\mathcal{H}(1+z)$, we can write

\begin{equation}
\label{DINT}
    \mathcal{D} = \frac{1}{2} \int_{0}^{z} \frac{a_{M}}{1+z^{\prime}} dz^{\prime}.
\end{equation}

Therefore we get the expression for primordial gravitational wave power spectrum in Einstein-Gauss-Bonnet gravity as,

\begin{eqnarray}
     \label{OMGWEGB}
    \Omega_{GW}(k) = \text{e}^{-2\mathcal{D}} \times \frac{k^{2}}{12H_{0}^{2}}r(k_{ref}) \mathcal{P}_{\zeta} (k_{ref}) \left( \frac{k}{k_{ref}}\right)^{n_{T}}  \left (\frac{\Omega_{m}}{\Omega_{\Lambda}}\right)^{2}  \left( \frac{g_{*}(T_{in})}{g_{*0}} \right)\nonumber \\ \left (\frac{g_{*s0}}{g_{*s}(T_{in})} \right)^{4/3} \left(\frac{3 j_{1}(k\tau_{0})}{k\tau_{0}}\right)^{2} T_{1}^{2}(x_{eq}) T^{2}_{2} (x_{R})  
\end{eqnarray}

Where we get the overall modification factor of $\text{e}^{-\mathcal{D}}$ in the standard expression of primordial GW power spectrum given by Eqn.(\ref{OMEGAGW12}).\\

For EGB models, compatible with GW170817, it was first noted in  \cite{Oikonomou:2022xoq}, that the factor $\mathcal{D}$, is vanishingly small, $\mathcal{O}(10^{-60})$, when the integral (\ref{DINT}) is calculated from present day up to the reheating era, $z \sim 10^{6}$. Thus, the factor of $\text{e}^{-2\mathcal{D}}$ becomes unity, and we can use the standard expression of power spectrum given by (\ref{OMEGAGW12}).

\section{Result and analysis of inflation, Reheating and GW}
\label{RES}
In the subsequent sections we present the results of the two inflationary models.

\subsection*{Natural Inflation}

\begin{figure}[!htb]
\centering
\includegraphics[scale=0.45]{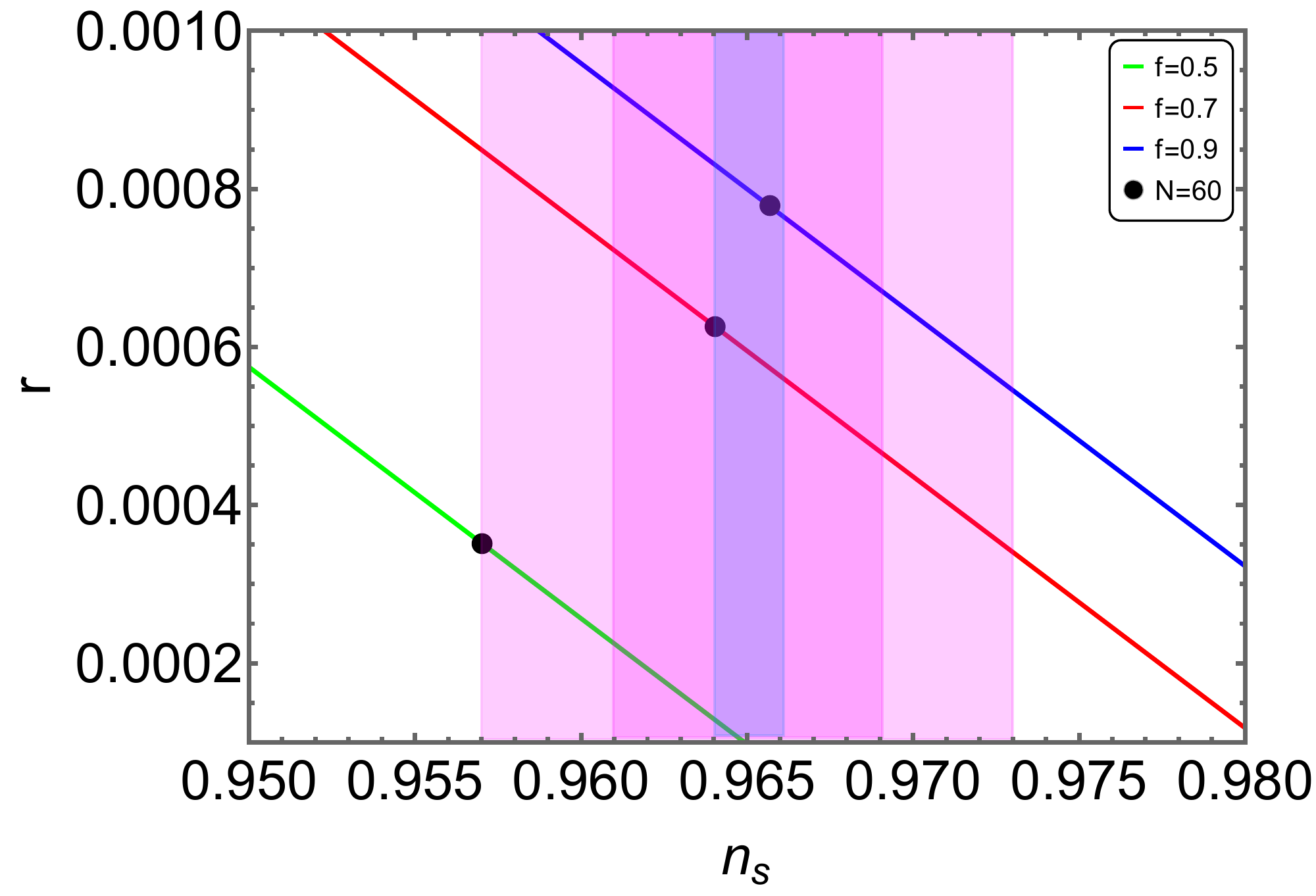}
\caption{Plots of $r$ and $n_s$, keeping $U=0.502$, for Natural inflation. The light pink shaded region corresponds to $2-\sigma$ and dark pink shaded region corresponds to $1-\sigma$ bounds on $n_s$ from {\it Planck'18}  \cite{Planck2018}. The deep blue shaded region corresponds to the $1-\sigma$ bounds of future CMB observations \cite{Euclid,PRISM} keeping the same central value.}
\label{rnsnat}
\end{figure}

Equipped with all the necessary equations for the estimation of the inflationary observables, we first calculate $r$ and $n_s$ for Natural inflation. The plots of $r$ against the $n_s$ are given in Fig.(\ref{rnsnat}), for  different values of $f$, keeping $U=0.502$. For the value of the constant $U$ we deviate a little from $U=0.5$, owing to the fact that at $U=0.5$, the effective potential $V_{eff} (\phi)$ becomes identically zero, irrespective of the potential $V (\phi)$ considered. 

\begin{figure}[!htb]
\centering
\includegraphics[scale=0.4]{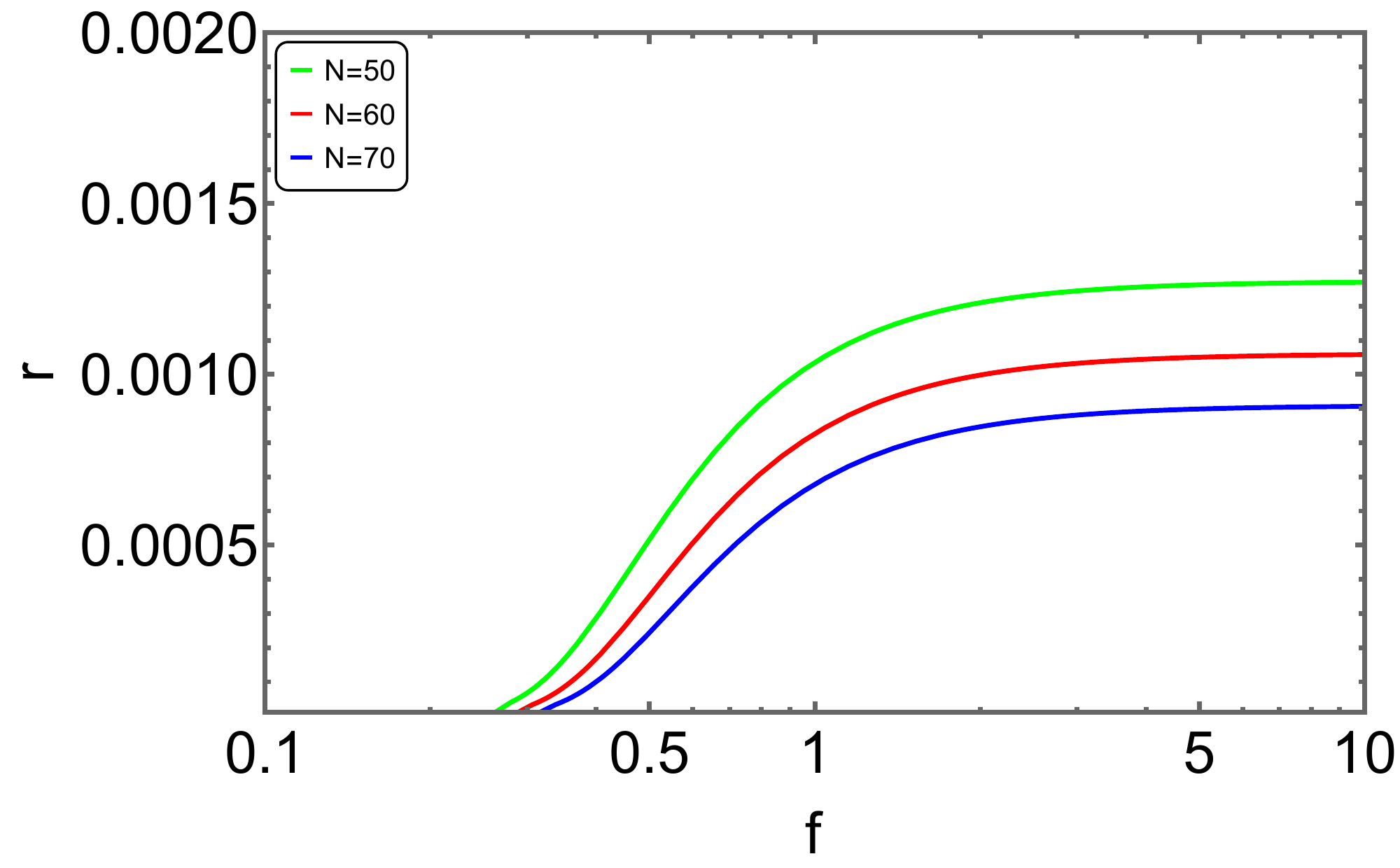}
\includegraphics[scale=0.4]{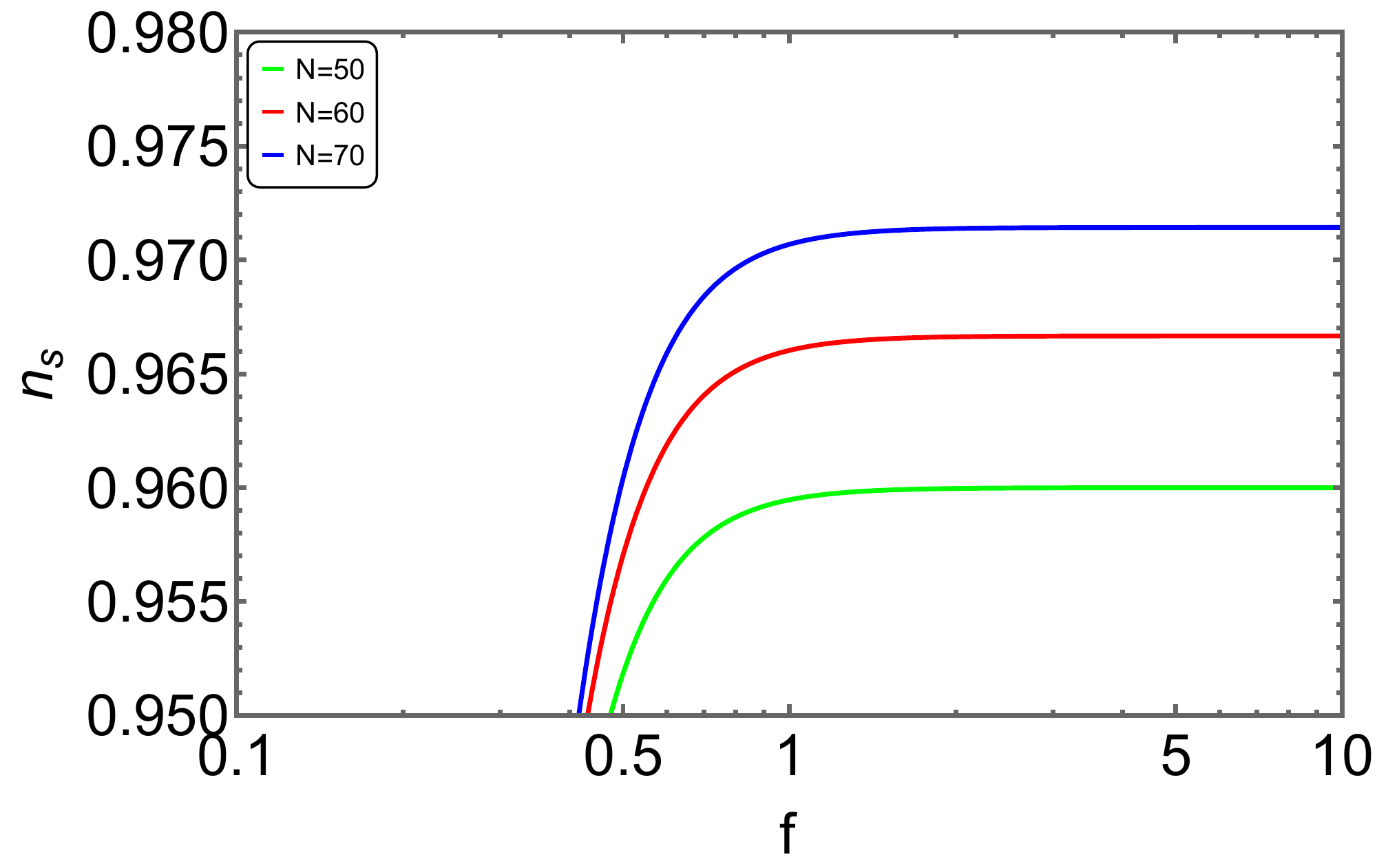}
\caption{Plots of $r$ and $n_s$ for Natural inflation against the $f$, for different numbers of e-folds, keeping $U=0.502$}
\label{r_ns_f_nat}
\end{figure}

Evolution of the $r$ and $n_s$ against $f$ is given in Fig.(\ref{r_ns_f_nat}).

\begin{center}
\begin{table*}[!ht]
\begin{center}
\begin{tabular}{|c|c|c|c|c|c|c|}
\hline
   & \multicolumn{2}{c|}{$f=0.5$} & \multicolumn{2}{c|}{$f=0.7$} & \multicolumn{2}{c|}{$f=0.9$} \\ \hline
$N$  & $n_s$             & $r$                                           & $n_s$             & $r$                                         & $n_s$             & $r$               \\ \hline
50 & 0.9518          & $5.1602 \times 10^{-4} $           & 0.9578         & $8.2367 \times 10^{-4}$       & 0.9591        & $9.8447 \times 10^{-4}$          \\ \hline
60 & 0.9570         & $3.5064 \times 10^{-4}$             & 0.9640          & $6.2514 \times 10^{-4}$       & 0.9657        & $7.7749 \times 10^{-4}$        \\ \hline
70 & 0.9604         & $2.4327 \times 10^{-4}$            & 0.9683          & $4.8698 \times 10^{-4}$        & 0.9703       & $6.3106 \times 10^{-4}$         \\ \hline
\end{tabular}
\end{center}
\caption{\textbf{Natural Inflation:} Values of the inflationary parameters $r$ and $n_{s}$ for different values of $f$ and number of e-folds $N$. The observables are in good agreement with the  $Planck'18$ \cite{Planck2018} }.
\label{tab1}
\end{table*}
\end{center}

\subsection*{Mutated Hilltop Inflation}
\begin{figure}[!htb]
\centering
\includegraphics[scale=0.45]{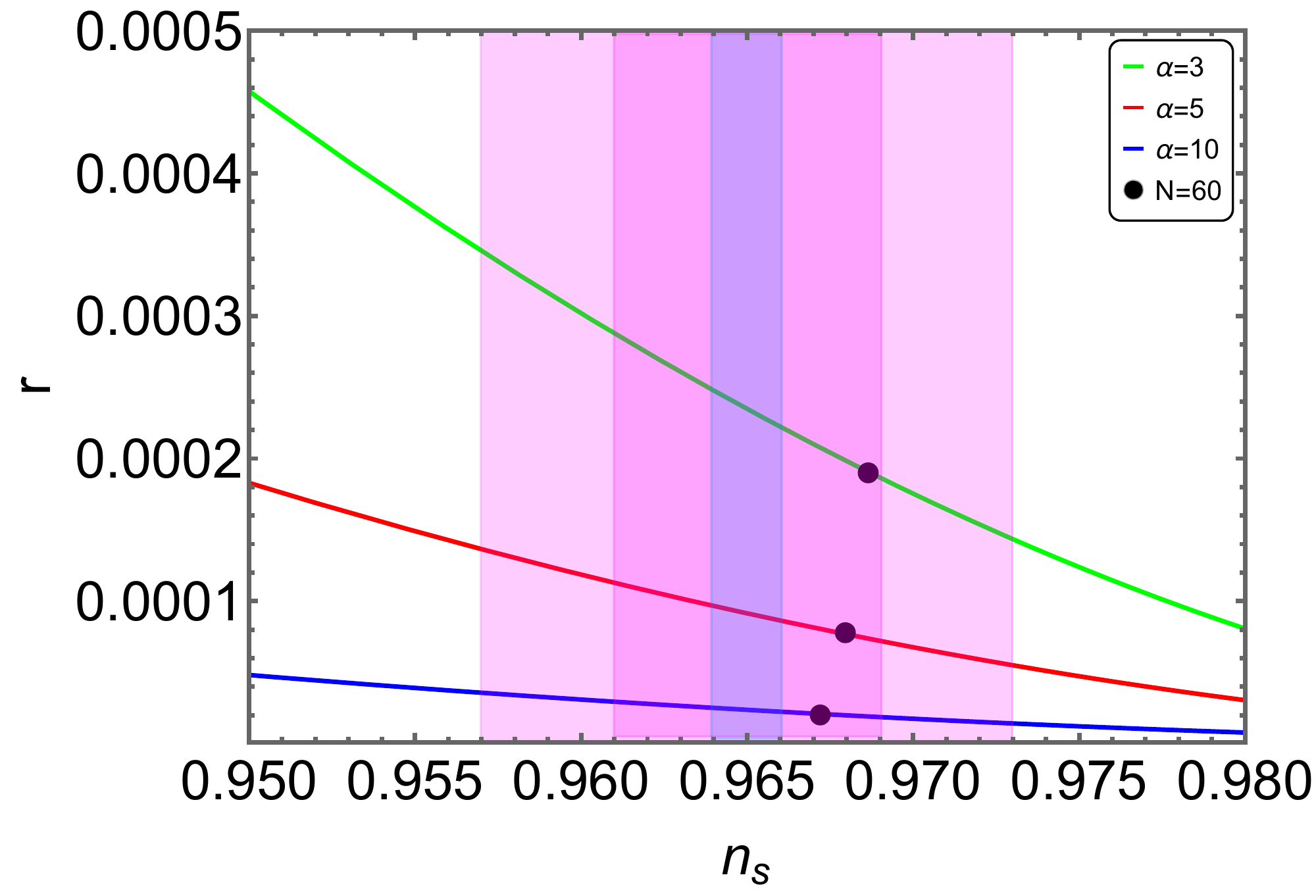}
\caption{Plots of $r$ and $n_s$, for $U=0.51$, for mutated hilltop inflation. The light pink shaded region corresponds to $2-\sigma$ and dark pink shaded region corresponds to $1-\sigma$ bounds on $n_s$ from {\it Planck'18} \cite{Planck2018}. The deep blue shaded region corresponds to the $1-\sigma$ bounds of future CMB observations \cite{Euclid,PRISM} keeping the same central value.   }
\label{r_ns_mut}
\end{figure}

\begin{figure}[!htb]
\centering
\includegraphics[scale=0.4]{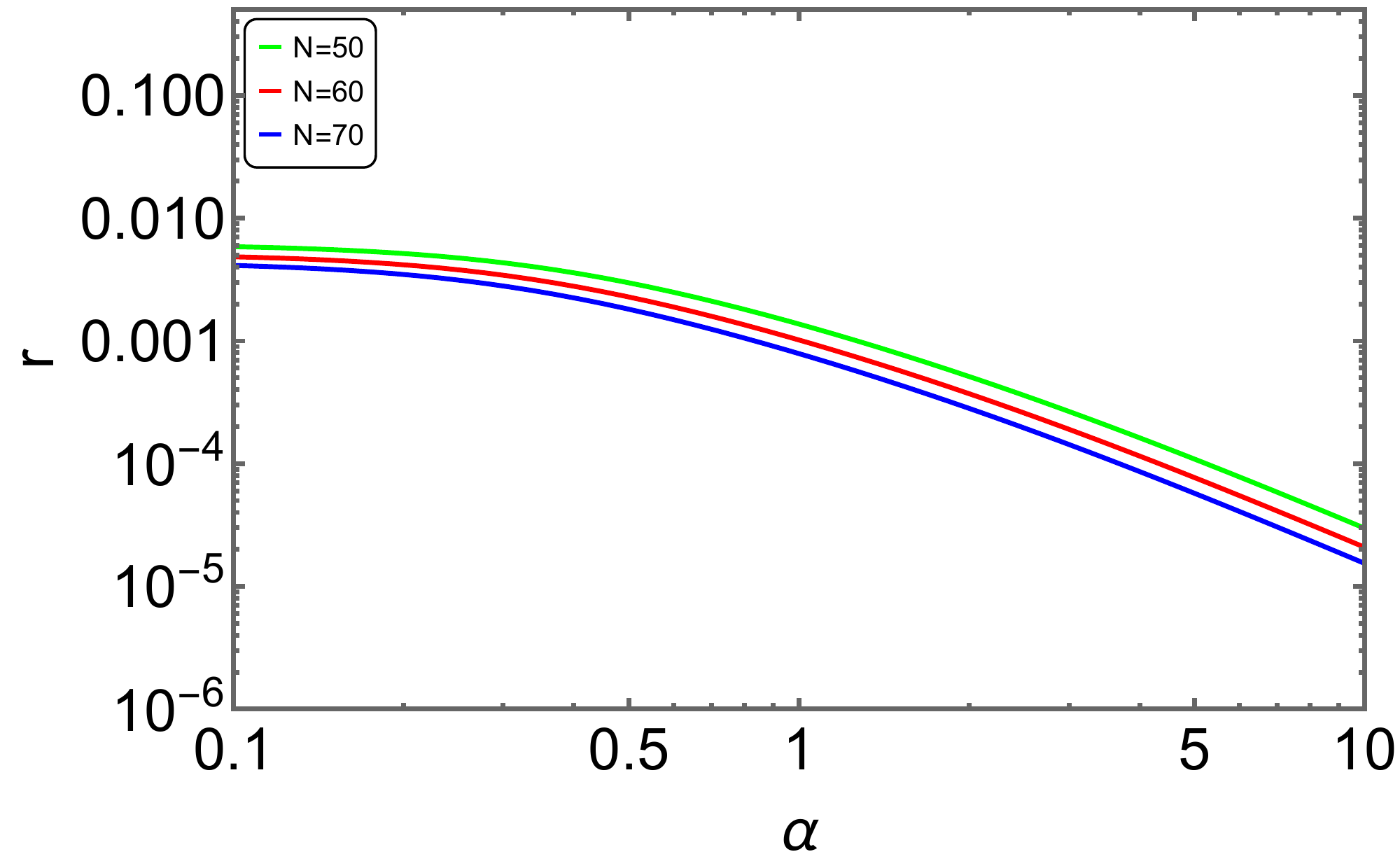}
\includegraphics[scale=0.4]{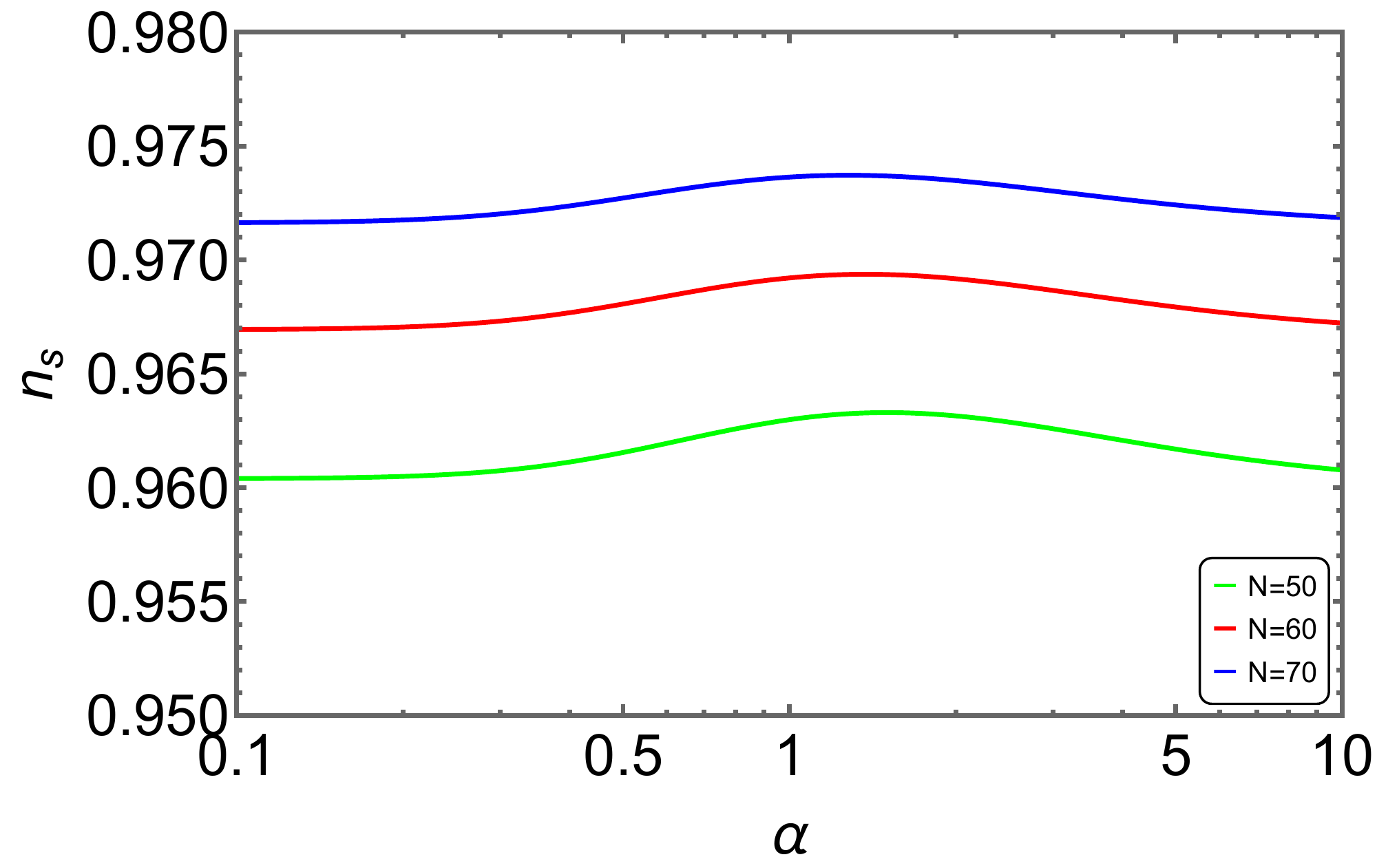}
\caption{Plots of $r$ and $n_s$, for mutated hilltop inflation as function of $\alpha$ for different number of e-folding keeping $U=0.51$ }
\label{r_ns_alpha}
\end{figure}

In this section we discuss the result of the inflationary dynamics for the mutated hilltop inflation model. Using the numerical method we compute the tensor to scalar ratio and spectral index which are drawn in fig. (\ref{r_ns_mut}) for three different values of $\alpha$, specific values of the parameter $\alpha$ are chosen to match with the {\it Planck'18} observational constraint \cite{Planck2018}, however it is possible to take the other values of the $\alpha$. Evolution of $r$ and $n_s$ with $\alpha$ are described in the fig. (\ref{r_ns_alpha})

\begin{center}
\begin{table*}[!ht]
\begin{center}
\begin{tabular}{|c|c|c|c|c|c|c|}
\hline
   & \multicolumn{2}{c|}{$\alpha=3$} & \multicolumn{2}{c|}{$\alpha=5$} & \multicolumn{2}{c|}{$\alpha=10$} \\ \hline
$N$  & $n_s$             & $r$                                           & $n_s$             & $r$                                         & $n_s$             & $r$               \\ \hline
50 & 0.9625          & $2.6620 \times 10^{-4} $           & 0.9616         & $1.0896 \times 10^{-4}$       & 0.9607        & $2.9727 \times 10^{-5}$          \\ \hline
60 & 0.9686         & $1.9048 \times 10^{-4}$             & 0.9679          & $7.7010 \times 10^{-5}$       & 0.9672        & $2.0797 \times 10^{-5}$        \\ \hline
70 & 0.9730         & $1.4322 \times 10^{-4}$            & 0.9724          & $5.7337 \times 10^{-5}$        & 0.9718       & $1.5364 \times 10^{-5}$         \\ \hline
\end{tabular}
\end{center}
\caption{{\bf Mutated Hilltop Inflation:} Values of the inflationary parameters $r$ and $n_{s}$ for different values of $\alpha$ and number of e-folds $N$. The observables are in good agreement with the  $Planck'18$ \cite{Planck2018} }.
\label{tab2}
\end{table*}
\end{center}
 
\begin{figure}[!htb]
\centering
\includegraphics[width=8cm,height=8cm]{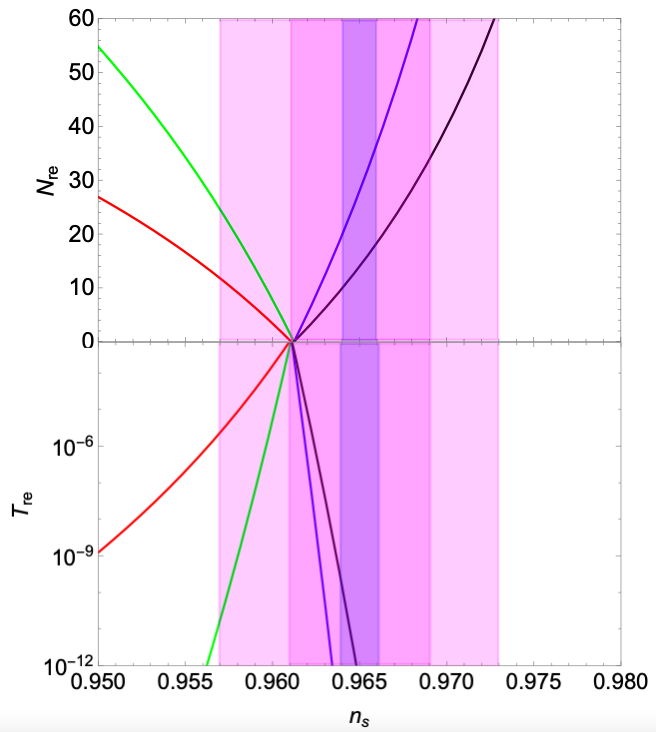}
\includegraphics[width=8cm,height=8cm]{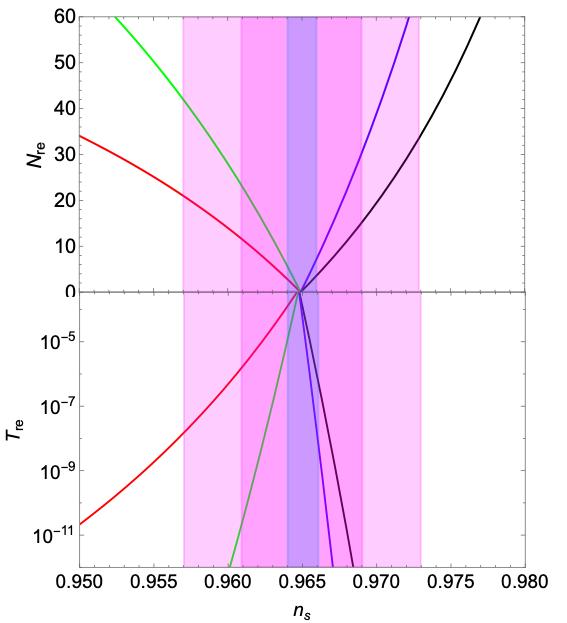}
\caption{Plots of $T_{re}$ and $N_{re}$ for different values of equation of state ($\omega_{re}$). {\bf Left Panel:} Plots of $T_{re}$ and $N_{re}$ for the Natural inflation, the red line corresponds to $\omega_{re}= \frac{-1}{3}$, green line corresponds to $\omega_{re}=0$, blue line stands for $\omega_{re}= \frac{2}{3}$ and black line is for $\omega_{re}= 1$. The light pink shaded region corresponds to $2-\sigma$ and dark pink shaded region corresponds to $1-\sigma$ bounds on $n_s$ from {\it Planck'18}\cite{Planck2018}. The deep blue shaded region corresponds to the $1-\sigma$ bounds of future CMB observations \cite{Euclid,PRISM} keeping the same central value. {\bf Right Panel:} Plots of $T_{re}$ and $N_{re}$ for the Mutated hilltop inflation, the red line corresponds to $\omega_{re}= \frac{-1}{3}$, green line corresponds to $\omega_{re}=0$, blue line stands for $\omega_{re}= \frac{2}{3}$ and black line is for $\omega_{re}= 1$. The light pink shaded region corresponds to $2-\sigma$ and dark pink shaded region corresponds to $1-\sigma$ bounds on $n_s$ from {\it Planck'18}\cite{Planck2018}. The deep blue shaded region corresponds to the $1-\sigma$ bounds of future CMB observations \cite{Euclid,PRISM} keeping the same central value.}
\label{reheating_plots}
\end{figure}

\newpage
\begin{center}
\bf{Reheating in Natural and Mutated Hilltop Inflation}    
\end{center}
In cold inflationary scenario, reheating is an inevitable epoch. In this section we try to analyse the thermalization temperature ($T_{re}$) and number of e-folding ($N_{re}$) during the reheating for a given inflationary model. From Eq. (\ref{re7}) and (\ref{re8}) it is straightforward to explore the reheating phase for the natural and mutated hilltop inlfation. Results of the reheating analysis are drawn in the fig. (\ref{reheating_plots}) 

\begin{figure}[!htb]
\centering
\includegraphics[width=8cm,height=6cm]{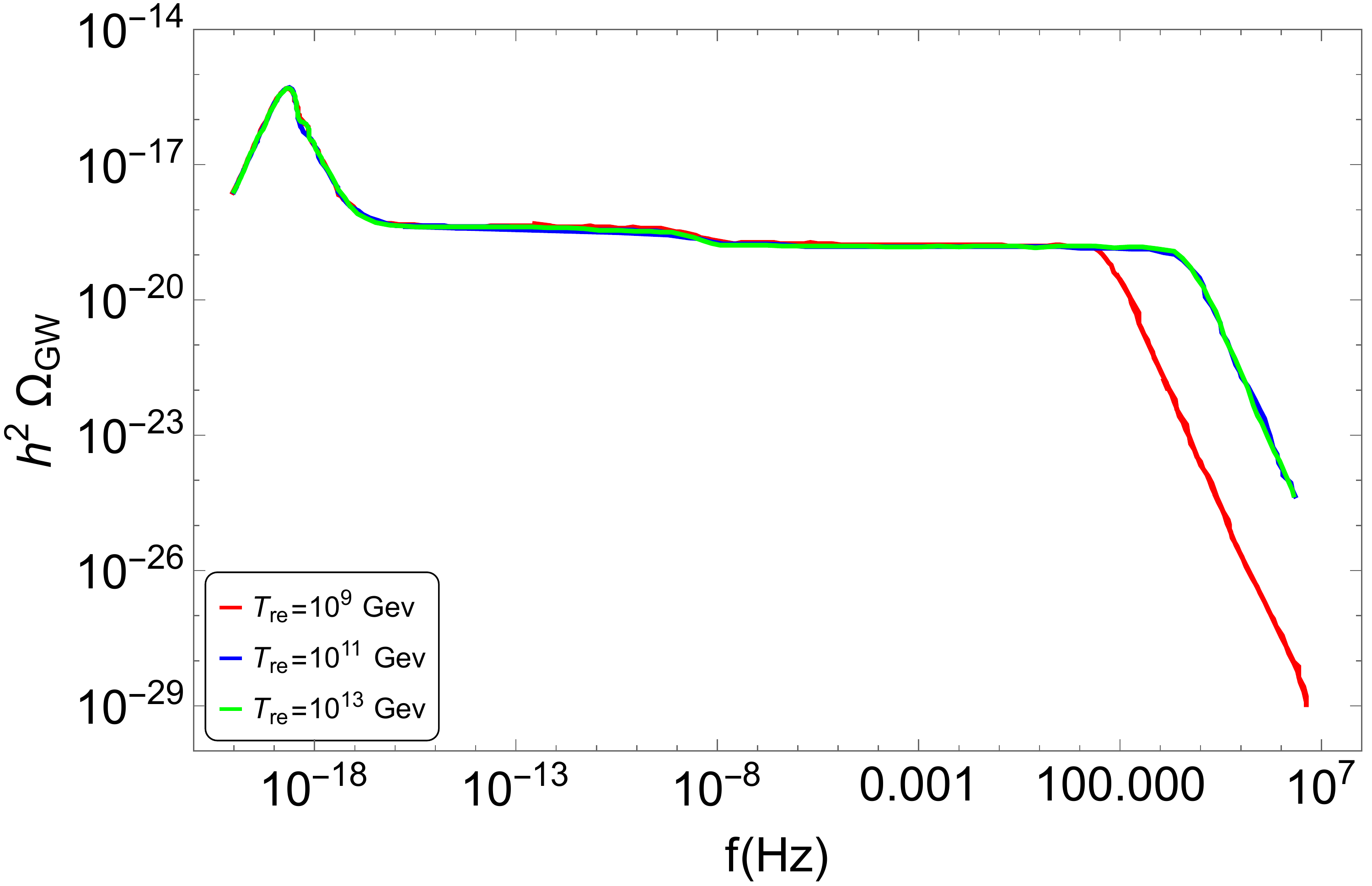}
\includegraphics[width=8cm,height=6cm]{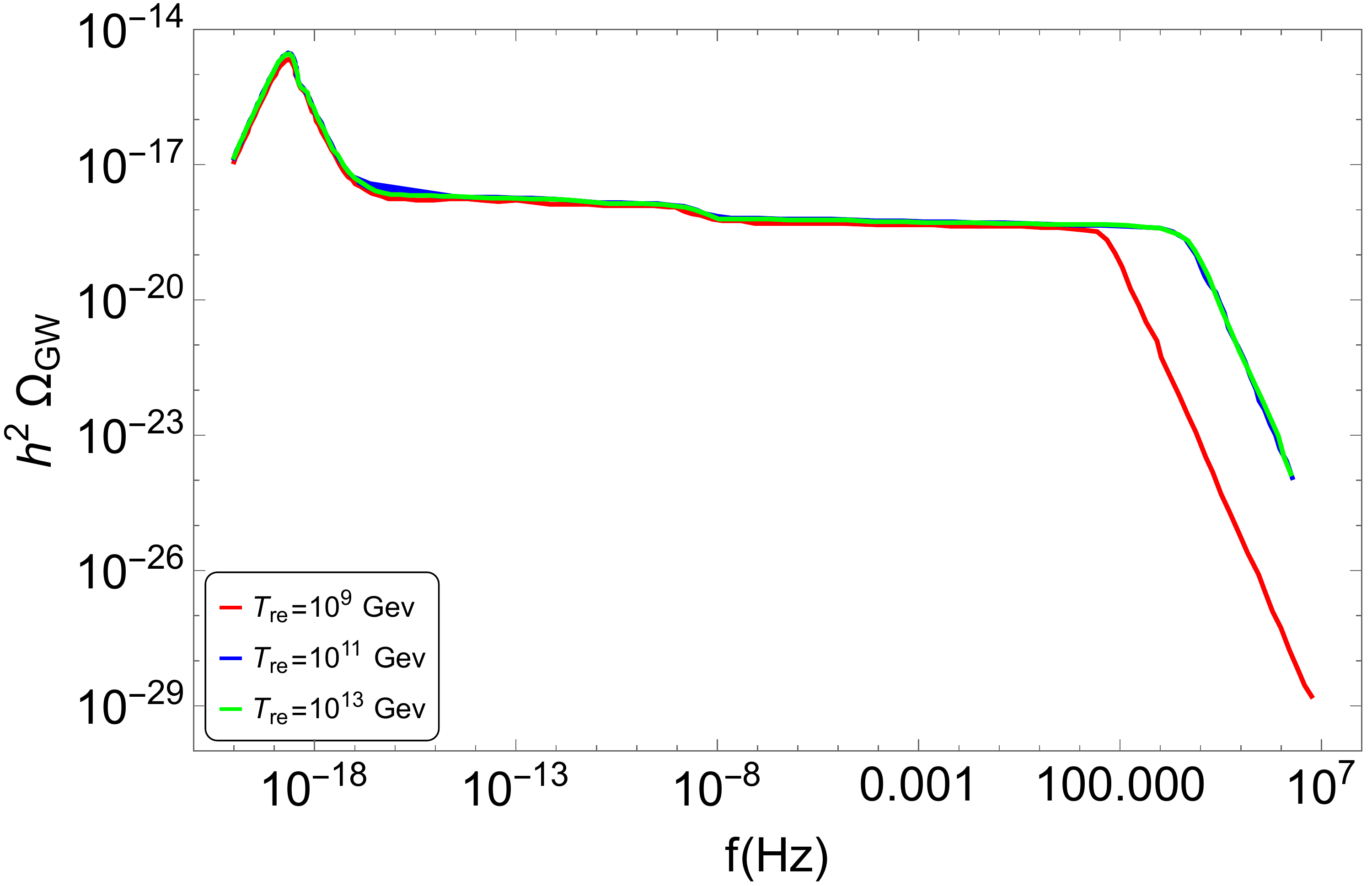}
\caption{Plots of $h^{2}$ scaled primordial gravitational wave energy spectrum for Mutated Hilltop potential (\textbf{Left}) and Natural Inflation (\textbf{Right}), for three different reheating temperatures. }
\label{GWPLOTS}
\end{figure}

\newpage
\begin{center}
\bf{Primordial Gravitaional Wave Energy Spectrum for Natural and Mutated Hilltop Inflation}   
\end{center}
In this section we will be calculating the primordial GW energy spectrum for Natural and Mutated Hilltop potentials. We use (\ref{OMEGAGW12}) to do the calculations as we have mentioned before that the extra factor of $\text{e}^{-2\mathcal{D}}$ is null in EGB-GW170817 compatible models \cite{Oikonomou:2022xoq}.\\
\noindent
The only three parameters that we need to input in Eqn.(\ref{OMEGAGW12}) from the potentials considered are $r$, $n_{T}$ and $T_{re}$. $T_{re}$ and $r$ have already been calculated in the previous sections. As for the value of the tensor spectral index $n_{T}$, in the case of Natural inflation, we calculated the GW spectrum for $n_{T} = -0.0122914$, at $f=0.9$ and $N=60$. For Mutated Hilltop inflation, we got $n_{T}=-0.000613379$, at $\alpha = 3$ and $N=60$. Figure (\ref{GWPLOTS}) shows the GW energy spectrum for Natural inflation and Mutated Hilltop inflation. As can be seen in from the plots, there is no significant enhancement in the GW spectrum for either of the inflationary potentials. This was expected, due to the fact that in both the cases, we got a red-tilted ($n_{T}< 0$) tensor spectral index. For a red-tilted tensor spectral index, it is rather difficult to obtain any kind of enhancement in the GW spectrum, due to the fact that that it is very sensitive on the value and sign of $n_{T}$. 

\section{Conclusion}
\label{conslusion}

In this paper, we studied  Natural inflation and Mutated Hilltop inflation in Einstein-Gauss-Bonnet gravity in the light of constraints put on the Gauss-Bonnet coupling $\xi (\phi)$ by \textit{GW170817} observation. As mentioned before, both of these models are well motivated, but Natural inflation faced a validity crisis due to observations constraints. So it was interesting to see how these models behaved in our framework. 

Natural inflation was studied for three different values of breaking scale $f(=0.5,0.7,0.9)$. In all three cases we found that the inflationary observables $r$ and $n_{s}$ are in good agreement with \textit{Planck'18} \cite{Planck2018}. The $r\;vs\;n_{s}$ plot for this analysis is given in Fig.(\ref{rnsnat}) and evolution of the $r,n_s$ with the breaking scale $f$ can be seen in the Fig.(\ref{r_ns_f_nat}). Values of inflationary observables for different number of e-folds are laid out in Table (\ref{tab1}). 

For Mutated Hilltop Inflation, we again took three different values of the parameter $\alpha(=3,5,10)$. For all the cases considered, the values of the inflationary parameters $r$ and $n_{s}$ were in agreement with the constraints of the \textit{Planck'18} \cite{Planck2018}. Again, the $r\;vs\;n_{s}$ plot is given in Fig.  (\ref{r_ns_mut}). Evolution of the $r$ and $n_s$ is shown in the Fig. (\ref{r_ns_alpha}) and the inflationary observables have been given in Table (\ref{tab2}). 

The analyses for reheating  was also carried out in this work. We studied the reheating dynamics in Natural inflation for $f=0.7$, and for Mutated Hilltop inflation, we took $\alpha = 5$. Although, we only considered one case each, for both the inflationary models, the reheating analyses can be done for all the cases. For both the models we found that the reheating parameters are in good agreement with the \textit{Planck'18}. Results for reheating are drawn in the Fig. (\ref{reheating_plots})

The analysis summarised above concludes that Natural inflation and Mutated Hilltop inflation in the Einstein-Gauss-Bonnet framework of inflation, are compatible with the \textit{Planck'18} and with the \textit{GW170817} constraints. 

Another important result that we got was the absence of any enhancement in the GW energy spectrum. 
In order to analyze the primordial gravitational wave energy spectrum, for both the models, we calculated the tensor spectral index. In the models that we considered here, we got a red-tilted ($n_{T}<0$) tensor spectral index. This was expected due to the fact that we are working in the effective potential construction in the Einstein-Gauss-Bonnet framework. Getting a blue tilted tensor spectral index is rather difficult to achieve, as we observed in the small field physical models that we considered in our work. Due to this fact, there is no enhancement in the GW spectrum for either of the potentials considered. The sign and value of $n_{T}$ play a fundamental role in (\ref{OMEGAGW12}), and we would expect such a spectrum for any model considered in the EGB framework, with a red-tilted tensor spectral index. 
The GW spectrum for both potentials are given in Figure (\ref{GWPLOTS}).

A few more studies that we plan to conduct in the Einstein-Gauss-Bonnet inflationary framework will be to see how the dynamics of Warm inflation\cite{44,44a, Correa:2022ngq} is modified, directly constrain the small field inflationary models in EGB framework, and also the production of Primordial Black holes \cite{Kawai:2021edk}. Finally, the long-lasting problem of low $l$ anomaly in the CMB power spectrum can be attributed to the modification of gravity sector\cite{Mathews:2015daa, Gangopadhyay:2017vqi}, thus EGB can have imprints in these scales. We hope to come back to these questions soon.

\section{Acknowledgement}
\noindent
Authors would like to thank Imtiyaz Ahmad Bhat and Mohit Kumar Sharma for the useful discussions. Work of M.R.G. is supported by Department of Science and Technology, Government of India under the Grant Agreement number IF18-PH-228 (INSPIRE Faculty Award) and by Science and Engineering Research Board(SERB), Department of Science and Technology(DST), Government of India under the Grant Agreement number  CRG/2022/004120 (Core Research Grant).
 H.A.K. would like to thank M.Sami and Centre for Cosmology and Science Popularization, SGT University for providing the facilities during the course of this work.

\end{document}